\documentclass[twoside]{article}
\usepackage{fleqn,espcrc2}

\usepackage{amssymb}

\newcommand{\lsim}{\stackrel{<}{_\sim}}
\newcommand{\cO}{{\cal O}}
\newcommand{\ket}{\,\rangle}
\newcommand{\bra}{\langle \,}

\newcommand{\es}{\varepsilon^{\phantom{a}}_{\mathrm{S}}}
\newcommand{\ep}{\varepsilon^{\phantom{a}}_{\mathrm{P}}}
\newcommand{\ppa}{\lambda_1^{\mathrm{PP}}}
\newcommand{\ppb}{\lambda_2^{\mathrm{PP}}}
\newcommand{\ppc}{\lambda_3^{\mathrm{PP}}}
\newcommand{\ssa}{\lambda_1^{\mathrm{SS}}}
\newcommand{\ssb}{\lambda_2^{\mathrm{SS}}}
\newcommand{\ssc}{\lambda_3^{\mathrm{SS}}}
\newcommand{\spa}{\lambda_1^{\mathrm{SP}}}
\newcommand{\spb}{\lambda_2^{\mathrm{SP}}}

\newcommand{\lam}{\lambda}

\title{One-loop Renormalization of Resonance Chiral Theory \\
with Scalar and Pseudoscalar Resonances
\thanks{Talk given at
the 12th International Quantum Chromodynamics Conference,
4--8th July (2005), Montpellier (France). IFIC/05-50; FTUV/05-1005 report.}}

\author{I.~Rosell \address{Departament de F\'{\i}sica Te\`orica, IFIC,
CSIC-Universitat de Val\`encia, \\ Apt. Correus 22085, E-46071 Val\`encia,
Spain}} 
       
\begin{document}

\begin{abstract}
The divergent part of the generating functional of the Resonance Chiral Theory is evaluated up to one loop when one multiplet of scalar and pseudoscalar resonances are included and interaction terms which couple up to two resonances are considered. Hence we obtain the renormalization of the couplings of the initial Lagrangian and, moreover, the complete list of operators that make this theory finite, at this order.
\vspace{1pc}
\end{abstract}

\maketitle

\section{Introduction}

With the current experimental and theoretical information it is known that Quantum Chromodynamics (QCD) describes correctly the hadronic physics. Because of the running of $\alpha_s$, one is not able to apply directly this procedure at low energies, where confinement prevents the use of the perturbative QCD --non-perturbative QCD is then needed--.

An alternative approach consists in an effective theory description in terms of the suitable degrees of freedom, which has been very successful in different branches of Particle Physics \cite{Pich,Manohar}. The main feature of this approach is the use of an effective Lagrangian with local operators involving only light particles, where the heavy particles have been integrated out. The high-energy dynamics is only kept in the low-energy couplings and through the symmetries of the full theory. 

At very low energies Chiral Perturbation Theory ($\chi$PT) is the effective theory of QCD, as dictated by the spontaneous breaking of the chiral symmetry, which allows the description of the hadronic regime in terms of the pseudo-Goldstone bosons, the pion multiplet, with a perturbative expansion in powers of the soft external momenta and masses of the pseudoscalar mesons \cite{ChPT1,ChPT2}. 

We are interested in QCD at energies between the $\rho$ mass and $2$ GeV, where the absence of a mass gap and the abundance of resonances make the effective theory approach more involved. Moreover there is no natural expansion parameter, since the use of the chiral counting is no longer valid at these energies.

Large-$N_C$ QCD provides an adequate framework to fulfill this aim. In this limit the Green Functions are described by the tree diagrams of an effective Lagrangian with local operators and an infinite number of meson fields, being higher corrections described by loops \cite{largeNC1,largeNC2}. 

Our approach involves using the Resonance Chiral Theory (R$\chi$T) \cite{RChTa,RChTb}, which describes QCD at intermediate energies ($M_\rho \lsim E \lsim 2 \,\mathrm{GeV}$) in terms of scalar, pseudoscalar, vector and axial-vector resonances besides the pseudo-Goldstone bosons. Although R$\chi$T follows the $1/N_C$ expansion, a model dependence appears when we consider only a finite number of resonances \cite{juanjo}. Actually this approximation is supported by the phenomenology and by the assumption that heavier resonances are suppressed by their masses.

Another remark is needed to understand the effective approach and particularly R$\chi$T: one is not working with an effective theory of QCD until the matching is considered \cite{Pich}. In this case, the matching at very low energies supports the use of the chiral symmetry in order to construct the Lagrangian and gives the leading contributions to the low energy couplings (LEC's) of $\chi$PT \cite{RChTa,portoles,jorge}; the matching at very high energies, via the operator product expansion and the Brodsky-Lepage conditions for the form factors, constraints the R$\chi$T couplings \cite{portoles,BrodskyLepage}.

Quantum loops in the R$\chi$T are necessary to improve the predictions and to get a better knowledge of non-perturbative QCD. At the scale of energies available experimentally nowadays, the importance of non-perturbative QCD to distinguish New Physics effects is obvious.

The aim of this work is to make a first step in the renormalization of the R$\chi$T \cite{treball}: the divergent part of the one-loop generating functional is evaluated when one multiplet of scalar and pseudoscalar resonances are included, and we allow for operators which couple up to two resonances. We obtain the renormalization of the couplings and the complete list of operators that make this theory finite at this order.

\section{The Resonance Effective Theory with scalar and pseudoscalar resonances}

Due to the large-$N_C$ limit \cite{largeNC1,largeNC2}, $U(3)$ multiplets for the spectrum are considered, while we prefer $SU(3)$ external currents as we are not interested in anomaly related issues. We allow for operators that contain pseudo-Goldstone bosons and states from the first multiplet of scalar and pseudoscalar resonances. 

Although in the initial R$\chi$T Lagrangian \cite{RChTa} only interaction terms with one resonance were included, we think that it is more convenient to consider here operators which couple up to two resonances, like in the kinetic pieces. Furthermore in a previous work \cite{VFF} it was conjectured that these new terms with more resonances are needed to keep the good short-distance behaviour at one loop; though this statement was not proved, it was observed that these terms eased the bad high-energy behaviour at tree level of some form factors with resonances in the final legs, two facts that seem to be related. In any case the requirement of the smooth behaviour of these form factors is an open question \cite{heretic}. 

As mentioned in the introduction, the short-distance properties of the underlying QCD must be implemented in the effective Lagrangian. This procedure establishes constraints among its couplings. Moreover, by considering the results of the Green functions of QCD through the operator product expansion (OPE), it turns out that resonance interactions with large number of derivatives are excluded. Therefore it seems natural to consider only operators with the minimum number of derivatives in the leading Lagrangian, an approximation which is corroborated by the phenomenology.

With all these ingredients our Lagrangian reads:
\begin{eqnarray}\label{Lagrangian}
\mathcal{L}_{\mathrm{R}\chi\mathrm{T}}(\mathrm{S},\mathrm{P})&=&\mathcal{L}^{(2)}_{\chi}\,+\,\mathcal{L}_{\mathrm{kin}}(\mathrm{S},\mathrm{P}) \,+\,  \mathcal{L}_{2}(\mathrm{S})\nonumber\\
&&+\,  \mathcal{L}_{2}(\mathrm{P})\,+\, \mathcal{L}_{2}(\mathrm{S},\mathrm{P}) \, ,
\end{eqnarray}
where the first piece is the $\cO(p^2)$ $\chi$PT Lagrangian,
\begin{equation}
\mathcal{L}_{\chi}^{(2)}\,=\,\frac{F^2}{4} \bra u_\mu u^\mu \, + \, \chi_+ \ket \,,
\end{equation}
where, as it is usual, $<...>$ is short for the trace in the flavour space and the other terms in Eq.~(\ref{Lagrangian}) introduce the terms with scalars and pseudoscalars, which have been split into the kinetic part,
\begin{equation}\label{kinetic}
\mathcal{L}_{\mathrm{kin}}  \,=\, \frac{1}{2} \sum_{R = S,P} \bra \nabla^\mu R \,\nabla_\mu R \,-\, M_R^2\, R^2 \ket \,, 
\end{equation}
and the $\cO(p^2)$ interactions linear and bilinear in the scalar and pseudoscalar fields \cite{jorge,treball,gerhard},
\begin{eqnarray}
\mathcal{L}_{2}(\mathrm{S})\!\!\! &= & \!\!\!  c_d \bra S \, u_\mu u^\mu \ket  + c_m \bra S \,\chi_+ \ket  \nonumber \\ 
\!\!\!& & \!\!\!+ \ssa \bra S^2 \,u^\mu u_\mu \ket 
+ \ssb \bra S u_\mu S u^\mu \ket \nonumber \\
\!\!\!&& \!\!\! + \ssc \bra S^2\, \chi_+ \ket \,,\\ \nonumber \\
\mathcal{L}_{2}(\mathrm{P})\!\!\!&=&\!\!\!i\,d_m \bra P \,\chi_- \ket  
+ \ppa \bra P^2 \,u^\mu u_\mu \ket \nonumber \\
\!\!\!& & \!\!\!+ \ppb \bra P u_\mu P u^\mu \ket  + \ppc \bra P^2\, \chi_+ \ket \,, \\ \nonumber \\
\mathcal{L}_{2}(\mathrm{S},\mathrm{P})\!\!\!& = & \!\!\!\spa\bra \{ \nabla_\mu S, P \} u^\mu \ket \nonumber \\
\!\!\!&& \!\!\!+\,i \spb \bra \{ S, P \} \chi_- \ket \, . 
\end{eqnarray}
The notation of Ref.\cite{RChTa,RChTb} is followed.

Note that as our Lagrangian satisfies the $N_C$ counting rules for an effective theory with $U(3)$ multiplets, only operators that have one trace in the flavour space are considered.

\section{Divergent part of the one-loop generating functional}

In order to evaluate the divergent part of the one-loop generating functional, an expansion around the classical solutions is made, in the spirit of the background field method \cite{background}. In our case, as we have pseudoscalar Goldstones, scalar and pseudoscalar resonances, one defines the quantum fluctuations as:
\begin{eqnarray} 
u_R\,=\,u_{cl}\,e^{i \Delta / 2} \, , \qquad\,\,\,\,\,
S\,=\,S_{cl}\,+\,\frac{1}{\sqrt{2}}{\es} \,, \nonumber \\
u_L\,=\,u_{cl}^\dagger  \, e^{-i\Delta / 2} \,, \qquad
P\,=\,P_{cl}\,+\,\frac{1}{\sqrt{2}}{\ep}\,,
\end{eqnarray}
with 
\begin{equation} \label{eq:fluc2}
\Delta \,=\,\Delta_i \lam_i / F \,, \quad 
\es\,=\,{\es}_{i}\,\lam_i\,, \quad
\ep\,=\,{\ep}_{i}\,\lam_i\,.
\end{equation}
In the following we will drop the subindex `$cl$' and all the fields will be understood to be classical. 

Inserting this expansion into the Lagrangian of Eq.~(\ref{Lagrangian}), and retaining terms quadratic in the quantum fields, we obtain the second-order fluctuation Lagrangian, which, after considering the equations of motion, can be written as:
\begin{equation}
\Delta {\cal L}_{\mathrm{R} \chi \mathrm{T}} \, = \, 
- \, \frac{1}{2} \, \eta \, \left( \, \Sigma_{\mu} \, \Sigma^{\mu} \, + \, 
\Lambda \, \right) \, \eta^{\top} \; , 
\end{equation}
where $\eta$ collects the fluctuations
, $\eta=\left(\Delta_i,{\es}_j,{\ep}_k\right)$, $i,j,k = 0,...,8$
, $\eta^{\top}$ is its transposed and $\Sigma_\mu$ and $\Lambda$ are $27 \times 27$ matrices \cite{treball}.

With the second-order fluctuation Lagrangian, and using the heat kernel techniques \cite{treball,background}, one is able to identify the divergences of the one-loop generating functional, specified by the action
\begin{equation}
S_{1} \, = \, \frac{i}{2} \, \ln \, \mbox{det} \, 
\left( \, \Sigma_{\mu} \, \Sigma^{\mu} \, + \, \Lambda \, \right) \; .
\end{equation}
Dimensional regularization is used to renormalize this determinant. Employing then the Schwinger-DeWitt proper-time representation, the divergent part of the action $S_1$ is found to be 
\begin{equation}\label{eq:oneloop}
S_{1}^{\,\mathrm{div}}\!\!=\!\!\frac{-1}{(4\pi)^2(D-4)} \! \int \!\!\mathrm{d}^4x \, \bra \!\!  \frac{1}{12}  Y_{\mu\nu} Y^{\mu\nu}  +   \frac{1}{2} \Lambda^2  \ket \,  , 
\end{equation}
where $Y_{\mu \nu}$ denotes the field strength tensor of $Y_{\mu}$,
\begin{equation}
Y_{\mu \nu} \,=\, \partial_{ \mu} Y_{\nu} - \partial_{\nu} Y_{\mu} + [Y_{\mu}, Y_{\nu}] \, ,
\end{equation}
where $Y_\mu$ is defined through the splitting
\begin{equation}
\left( \Sigma_\mu \right)_{ij} \,=\,\delta_{ij} \, \partial_\mu\, + \, \left( Y_{\mu} \right)_{ij} \,.
\end{equation}

\section{Results and conclusions}
In the renormalization of effective field theories by means of dimensional regularization, $S_1^{\, \mathrm{div}}$ might be absorbed by the redefinition of the couplings of the next-to-leading Lagrangian, getting thus a finite quantum field theory at this order. In our case we get the following subleading Lagrangian,
\begin{equation} \label{NLO}
{\cal L}_{1} \, = \,   \sum_ {i=1}^{18}  \alpha_i \, 
{\cal O}_i \, +  \sum_{i=1}^{68}  \beta_i^R \, {\cal O}_i^R \, +  
\sum_{i=1}^{383}  \beta_i^{RR} \, {\cal O}_i^{RR}  .
\end{equation}
The ${\cal O}_i$, ${\cal O}_i^{R}$ and ${\cal O}_i^{RR}$ operators involve zero, one and two resonance fields respectively. The couplings in the bare Lagrangian $ {\cal L}_{1}$ read:
\begin{eqnarray} \label{running}
\alpha_i & \!\!= & \!\! \mu^{D-4}  \left(  \alpha_i^r(\mu)  +  
\frac{1}{(4\pi)^2}  \frac{1}{D-4}  \gamma_i  \right)  , \nonumber \\
\beta_i^R & \!\! = & \!\! \mu^{D-4}  \left( \beta_i^{R,r}(\mu)  +  
\frac{1}{(4\pi)^2}  \frac{1}{D-4}  \gamma_i^R  \right)  , \nonumber \\
\beta_i^{RR} & \!\! = & \!\! \mu^{D-4}  \left( \beta_i^{RR,r}(\mu)  +  
\frac{1}{(4\pi)^2}  \frac{1}{D-4}  \gamma_i^{RR}  \right)  , \nonumber\\
\end{eqnarray}
where  $\gamma_i$, $\gamma_i^R$ and $\gamma_i^{RR}$ are the divergent coefficients given by $S_1^{\mathrm{div}}$ that constitute the $\beta$-function of our Lagrangian. The list of these operators, needed for the renormalization, and the values of $\gamma_i$, $\gamma_i^R$ and $\gamma_i^{RR}$, which fix the running of the couplings, appear partially in \cite{treball} and fully in \texttt{http://ific.uv.es/quiral/rt1loop.html}.

Some remarks are convenient here in order to understand these results:
\begin{enumerate}
\item To be consistent with the initial election, we have considered only the operators $\mathcal{O}_i$ in equation (\ref{NLO}) which couple up to two resonances. Moreover, as explained in Ref.~\cite{treball}, a cut in the number of resonances is needed in the procedure to perform the functional integration.
\item As the Lagrangian of Eq.~(\ref{Lagrangian}) only considers terms with the minimum number of derivatives, $\mathcal{O}(p^2)$, the operators of $\mathcal{L}_1$ are constructed by resonances and chiral tensors up to $\mathcal{O}(p^4)$.
\item Although the number of operators in (\ref{NLO}) is very large, one should keep in mind that we are studying only the divergent part of the couplings. We expect that most of the finite part of them must vanish in order to recover a good short-distance behaviour.
\end{enumerate}

Our result provides the running of the $\alpha_i$, $\beta_i^R$ and $\beta_i^{RR}$ couplings through the renormalization group equations (RGE). From Eq.~(\ref{running}) we get~:
\begin{equation} \label{eq:rgeq}
\mu \, \frac{d}{d \mu} \, \alpha_i^r(\mu) \, = \, 
- \, \frac{\gamma_i}{16 \,\pi^2} \; , 
\end{equation}
and, analogously, for $\beta_i^R$ and $\beta_i^{RR}$. These results can be potentially useful if we are interested in the phenomenological evaluation of the resonance couplings at this order. Within this issue it is interesting to take a closer look to the running of the resonance couplings in the original R$\chi$T Lagrangian \cite{RChTa}, namely, $c_d^r(\mu)$, $c_m^r(\mu)$ and $d_m^r(\mu)$, once the large-$N_C$ relations of the couplings are used \cite{treball}. Thus we predict no evolution for $c_m^r(\mu)$ and $d_m^r(\mu)$, while unfortunately we cannot conclude anything about $c_d^r(\mu)$, as there are no known constraints on $\lambda_1^{SS}$ and $\lambda_2^{SS}$.
\vspace{0.4cm}

\noindent {\bf Acknowledgments}  \\
I wish to thank S.~Narison for the organization of the 12th International QCD Conference and J.~Portol\'es and P.D.~Ruiz-Femen\' \i a for their helpful comments. My work is supported by a FPU scholarship of the Spanish MEC. This work has been supported in part by the EU HPRN-CT2002-00311 (EURIDICE), by MEC (Spain) under grant FPA2004-00996 and by Generalitat Valenciana under grants GRUPOS03/013, GV04B-594 and GV05/015.

\end{document}